\documentclass[10pt,conference]{IEEEtran}

\usepackage{research5}
\usepackage{psfrag}
\usepackage{epsfig}
\usepackage{graphicx}
\usepackage{multicol}
\usepackage{cite} 

\newtheorem{theorem}{Theorem}
\newtheorem{lemma}{Lemma}

\newtheorem{proposition}{Proposition}

\newcommand{\tr}{\mathsf{tr}}

\title{Optimal Throughput for Covert Communication Over a Classical-Quantum Channel}

\author{\authorblockN{Ligong Wang} \authorblockA{ETIS, ENSEA - Universit\'e de Cergy-Pontoise - CNRS\\Cergy-Pontoise, France\\ \texttt{ligong.wang@ensea.fr}}}

\begin{document}

\maketitle 
\begin{abstract}
This paper considers the problem of communication over a memoryless classical-quantum wiretap channel subject to the constraint that the eavesdropper on the channel should not be able to learn with high confidence whether the legitimate parties are using the channel to communicate or not. Specifically, the relative entropy between the output quantum states at the eavesdropper when a codeword is transmitted and when no input is provided must be sufficiently small. Extending earlier works, this paper proves the ``square-root law'' for a broad class of classical-quantum channels: the maximum amount of information that can be reliably and covertly transmitted over $n$ uses of such a channel scales like $\sqrt{n}$. The scaling constant is also determined.
\end{abstract}


\section{Introduction}\label{sec:intro}

Covert communication \cite{bloch16}, sometimes also called deniable communication \cite{chebakshijaggi13}, stealth \cite{houkramer14}, or communication with low probability of detection \cite{bashgoekeltowsley13,wangwornellzheng15,wangwornellzheng16}, refers to secret-communication scenarios where the legitimate parties must keep the eavesdropper from discovering the fact that they are using the channel to communicate. 

Many recent works study covert communication using information-theoretic methods. Korzhik et al. \cite{korzhik05} and Bash et al. \cite{bashgoekeltowsley13} observe that the amount of information that can be reliably and covertly communicated over an additive white Gaussian noise (AWGN) channel scales like the square root of the number of times the channel is used. A similar result has been known in steganography as the ``square-root law''; see \cite{fridrich09} and references therein. Che et al. \cite{chebakshijaggi13} prove that the square-root law can be achieved over the binary symmetric channel. They also show that, in some regimes, this can be done without employing a secret key. Wang et al. \cite{wangwornellzheng15,wangwornellzheng16} subsequently prove both achievability (with secret key) and converse of the square-root law for a broad class of discrete memoryless channels (DMCs), and also derive the exact scaling constant for DMCs as well as for AWGN channels. Bloch \cite{bloch16} studies covert communication from a resolvability perspective and derives, in addition to the scaling law of the throughput, the minimum length of secret key that is required.

Covert communication over various types of quantum channels has also been studied. For example, Bash et al. \cite{bash15} study covert quantum optical communication. Sheikholeslami et al.~\cite{sheikholeslami16} prove the achievability part of the square-root law for more general classical-quantum channels: the inputs of such channels are drawn from sets of (classical) symbols, and the outputs are quantum states. They leave the converse of the square-root law for such channels open. 

The current paper extends the work of \cite{sheikholeslami16} to determine the optimal throughput for covert communication over a classical-quantum channel up to the dominant term. Our main result, Theorem~\ref{thm:main}, establishes the square-root law for a broad class of classical-quantum channels, and also determines the scaling constant. Theorem~\ref{thm:main} can be seen as a counterpart in the classical-quantum case of \cite[Theorem 2]{wangwornellzheng16}, with one minor generalization: \cite{wangwornellzheng16} assumes that the eavesdropper observes the same channel outputs as the intended receiver, while the current paper assumes a wiretap channel structure, where the received states at the intended receiver and the eavesdropper may be different. 

The main new technial ingredient in this paper is a converse proof adapted from the classical one in \cite{wangwornellzheng16}.\footnote{Both the current paper and \cite{sheikholeslami16} consider only classical-quantum channels, for which entangled input states cannot be employed. Extending our converse to purely quantum channels (which allow entangled input states) may not be easy; see also Section~\ref{sec:conclusion}.} Our achievability proof largely follows \cite{sheikholeslami16}. Whereas \cite{sheikholeslami16} considers channels with only two distinct input symbols, we allow the input alphabet to be any finite set, which does not add significant complication to the problem. On the other hand, \cite{sheikholeslami16} also derives a bound on the length of the secret key required for covert communication, which is not done in the current paper.

The rest of this paper is arranged as follows. Section~\ref{sec:results} presents the channel model and the results. Section~\ref{sec:proof} proves the main theorem; the converse proof is given completely, while in the achievability proof we skip some less important steps, as our techniques there are similar to those in~\cite{sheikholeslami16}. We conclude the paper with some remarks in Section~\ref{sec:conclusion}.

\section{Channel Model and Results}\label{sec:results}

Consider a classical-quantum wiretap channel whose input alphabet is $\mathcal{X} = \{0,\ldots,k-1\}$. When a symbol $x\in\mathcal{X}$ is transmitted, the corresponding channel output at the intended receiver is a quantum state, described by a density operator $\sigma(x)$ acting on some Hilbert space $\mathbb{Y}$. The corresponding channel output at the eavesdropper is another quantum state $\rho(x)$ on some other Hilbert space $\mathbb{Z}$.\footnote{Once the two output states are given, their joint state on $\mathbb{Y}\otimes\mathbb{Z}$ is irrelevant to the problem considered in this paper, hence we make no further assumption on it. Note, however, that we do not assume that $\rho(x)$ is a purification of $\sigma(x)$, so the two states need not have the same eigenvalues.} We assume that both $\mathbb{Y}$ and $\mathbb{Z}$ have finite dimensions. The channel is assumed to be memoryless, so when a codeword $x^n = (x_1,\ldots,x_n)$ is transmitted over $n$ uses of the channel, the joint output states at the intended receiver and at the eavesdropper are described by the density operators $\sigma(x_1)\otimes\cdots\otimes\sigma(x_n)$ on $\mathbb{Y}^{\otimes n}$ and $\rho(x_1)\otimes\cdots\otimes\rho(x_n)$ on $\mathbb{Z}^{\otimes n}$, respectively. 

A deterministic channel code of blocklength\footnote{We use the usual terminology ``blocklength'' to refer to the total number of channel uses by a code. However, it should be kept in mind that, in the square-root case, the code can only be used once, because repeated transmissions will eventually be discovered by the eavesdropper.} $n$ for the message set $\mathcal{M}$ consists of an encoder $f\colon \mathcal{M}\to \mathcal{X}^n$, $m\mapsto x^n$, and a decoding positive-operator valued measure (POVM) on $\mathbb{Y}^{\otimes n}$; the outcome of the POVM is the decoded message. A random code of blocklength $n$ for $\mathcal{M}$ is described by a probability distribution on all deterministic codes of blocklength $n$ for $\mathcal{M}$. The average probability of error of a random code is the average probability that the output of the decoding POVM differs from the transmitted message $M$, where $M$ is uniformly chosen from $\mathcal{M}$, independently of the random choice of codebook.

Let $0$ be the ``off'' symbol in $\mathcal{X}$, which the transmitter always transmits when switched off. The covertness constraint is that the state observed by the eavesdropper should look sufficiently close to $\rho(0)^{\otimes n}$, which is the state it would observe if the transmitter were switched off. Specifically, as in \cite{wangwornellzheng16,sheikholeslami16}, our covertness constraint is expressed in terms of relative entropy
\begin{equation}\label{eq:LPD}
	D\left(\rho^n\left\| \rho(0)^{\otimes n} \right.\right) \le \delta
\end{equation}
for some positive $\delta$. Here $\rho^n$ denotes the joint output state at the eavesdropper over $n$ channel uses produced by the randomly chosen code and the uniformly chosen message, and the quantum relative entropy $D(A\| B)$ is defined, for any two density operators $A$ and $B$ acting on the same Hilbert space, as
\begin{equation}
	D(A\| B) = \begin{cases} \mathsf{tr} \left[ A \log A - A \log B\right],& \mathsf{supp}(A) \subseteq \mathsf{supp}(B),\\ \infty,& \textnormal{otherwise.}\end{cases}
\end{equation}
Obviously, if, for some $x\in\set{X}$, $\mathsf{supp}(\rho(x))\not\subseteq\mathsf{supp}(\rho(0))$, then using $x$ will result in the left-hand side of \eqref{eq:LPD} being infinite, and violating the covertness constraint for any $\delta$. Hence, for the purpose of the current work, we can discard all such input symbols. This allows us to assume, for the rest of this paper, that
\begin{equation}\label{eq:support0}
\mathsf{supp}(\rho(0)) = \mathbb{Z}.
\end{equation}

We are interested in characterizing the maximum amount of information that can be transmitted over a classical-quantum channel such that \eqref{eq:LPD} is satisfied, and that the probability of a decoding error can be made arbitrarily small at sufficiently large blocklengths. Like in the classical case \cite{wangwornellzheng16,bloch16}, this quantity can scale with $n$ in different ways. The key property that distinguishes between these cases is whether $\rho(0)$ can be written as a mixture of $\{\rho(1),\ldots,\rho(k-1)\}$ or not.

\subsection{Positive-Rate Case}

The following proposition characterizes the maximum achievable rate for covert communication over a classical-quantum channel.

\begin{proposition}\label{prp:rate}
The maximum rate at which information can be transmitted over the given channel with average error probability tending to zero as the blocklength tends to infinity and subject to the covertness constraint \eqref{eq:LPD}, in nats per channel use, is given by
\begin{equation}
	\max_{P\colon \sum_x P(x)\rho(x)=\rho(0)} \chi (P),
\end{equation}
where $\chi(P)$ is the Holevo information over the channel to the intended receiver \cite{holevo98}:
\begin{equation}
	\chi(P) = H \left(\sum_{x\in\mathcal{X}} P(x) \sigma(x) \right) - \sum_{x\in\mathcal{X}} P(x) H\left(\sigma(x)\right),
\end{equation}
with $H(\cdot)$ denoting the von Neumann entropy:
\begin{equation}
	H(A) = - \mathsf{tr}[A\log A].
\end{equation}
\end{proposition}

The proof of Proposition~\ref{prp:rate} is similar to the classical case \cite[Proposition 1]{wangwornellzheng16} and is omitted.

Proposition~\ref{prp:rate} implies that a positive rate is achievable for our task if, and only if, a distribution $P$ on $\mathcal{X}$ exists that satisfies the following conditions:
\begin{subequations}\label{eq:mixture}
\begin{IEEEeqnarray}{rCl}
&& \sum_{x} P(x) \rho(x)  =  \rho(0)\label{eq:mixture1} \\
&& \exists x \in\mathsf{supp}(P)\colon \sigma(x) \neq \sigma(0).
\end{IEEEeqnarray}
\end{subequations}

\subsection{Square-Root Case}

Now we assume that no $P$ satisfying \eqref{eq:mixture} can be found. In addition, we assume that
\begin{equation}\label{eq:suppsubset}
	\mathsf{supp} \left(\sigma(x)\right) \subseteq \mathsf{supp} \left(\sigma(0)\right)\quad \textnormal{for all }x\in\mathcal{X}.
\end{equation}
This case is the main focus of the current paper. Let $K_n(\delta,\epsilon)$ denote the maximum number of nats that can be sent over $n$ uses of this wiretap channel, with average probability of error at most $\epsilon$, and satisfying the covertness constraint \eqref{eq:LPD}. Further define
\begin{equation}\label{eq:defL}
	L \triangleq \lim_{\epsilon \downarrow 0} \liminf_{n\to\infty} \frac{K_n(\delta,\epsilon)}{\sqrt{n\delta}}.
\end{equation}
The main result of this paper is the following theorem, which can be seen as a generalization of \cite[Theorem 2]{wangwornellzheng16}.
\begin{theorem}\label{thm:main}
If no $P$ satisfying \eqref{eq:mixture} can be found, and if \eqref{eq:suppsubset} holds, then $L$ is finite and is given by
\begin{equation}\label{eq:main}
L = \max_{\tilde{P}\colon \tilde{P}(0) = 0} \frac{\sum_{x\in\mathcal{X}} \tilde{P}(x) D\left( \left. \sigma(x) \right\| \sigma(0) \right)}{\sqrt{\displaystyle\frac{1}{2} \eta\left(\left.\tilde\rho \right\| \rho(0)\right)}},
\end{equation}
where $\tilde{\rho}$ is determined by $\tilde{P}$ via
\begin{equation}
	\tilde{\rho} \triangleq \sum_{x\in\mathcal{X}} \tilde{P}(x) \rho(x),
\end{equation}
and where, for density operators $A$ and $B$,
\begin{equation}\label{eq:defeta}
\eta(A\|B) \triangleq \mathsf{tr}\left[ \int_0^\infty A (B+s)^{-1} A (B+s)^{-1} \d s \right] - 1.
\end{equation}
\end{theorem}

The proof of Theorem~\ref{thm:main} is given in Section~\ref{sec:proof}.

\subsection{Super-Square-Root Case}
Finally consider the case where no $P$ satisfying \eqref{eq:mixture} can be found, and where \eqref{eq:suppsubset} does not hold. In this case, no positive covert communication rate can be achieved, while $L$ as defined in \eqref{eq:defL} is infinite, hence the optimal throughput scales in a sub-linear but super-square-root way with the blocklength $n$. We do not further investigate this case in the current paper. The corresponding classical case is discussed in \cite[Appendix~G]{bloch16}.

\section{Proof of Theorem~\ref{thm:main}}\label{sec:proof}

In this section we prove Theorem~\ref{thm:main}. We divide the proof into the converse part and the achievability part.

\subsection{Converse}
Consider a random code that transmits $K_n$ nats of information over $n$ uses of the given channel with error probability not exceeding $\epsilon_n$. Let $P^n$ denote the distribution on $n$ input symbols induced by the random code and the uniformly drawn message, and let $\sigma^n$ denote the joint output state over $n$ channel uses at the intended receiver. By Fano's inequality and the data processing inequality, we have
\begin{equation}\label{eq:In}
	K_n (1-\epsilon_n) -1 \le \chi^n(P^n),
\end{equation}
where $\chi^n(P^n)$ is the Holevo information over $n$ uses of the channel to the intended receiver, computed according to $P^n$. Let $P_i$ and $\sigma_i$ denote the $i$th marginal distribution of $P^n$ and $i$th partial trace of $\sigma^n$, respectively. We have
\begin{IEEEeqnarray}{rCl}
	\chi^n(P^n) & = & H\left(\sigma^n\right) - \sum_{x^n\in\set{X}^{\otimes n}} P^n(x^n) H \left(\bigotimes_{i=1}^n \sigma(x_i)\right)\IEEEeqnarraynumspace \\	
	& = & H\left(\sigma^n\right) - \sum_{x^n\in\set{X}^{\otimes n}} P^n(x^n) \sum_{i=1}^n H\left(\sigma(x_i)\right)\\
	& = & H\left(\sigma^n\right) - \sum_{i=1}^n \sum_{x_i\in\mathcal{X}} P_i(x_i) H\left(\sigma(x_i)\right)\\
	& \le & \sum_{i=1}^n H\left(\sigma_i\right) - \sum_{i=1}^n \sum_{x_i\in\mathcal{X}} P_i(x_i) H\left(\sigma(x_i)\right)\\
	& = & \sum_{i=1}^n \chi(P_i).\label{eq:breaking}
\end{IEEEeqnarray}
Further, because Holevo information is concave in the input distribution (a consequence of the concavity of the von Neumann entropy), we can write
\begin{equation}\label{eq:concave}
	\sum_{i=1}^n \chi(P_i) \le n\, \chi\left(\overline{P}\right),
\end{equation}
where $\overline{P}$ is the average input distribution
\begin{equation}
	\overline{P} \triangleq \frac{1}{n} \sum_{i=1}^n P_i.
\end{equation}
Combining \eqref{eq:In}, \eqref{eq:breaking}, and \eqref{eq:concave} we obtain
\begin{equation}\label{eq:converse1}
	\frac{K_n(1-\epsilon_n)-1}{\sqrt{n\delta}} \le \sqrt{\frac{n}{\delta}} \,\chi\left(\overline{P}\right).
\end{equation}

We next consider the covertness criterion. Let $\rho_i$ denote the $i$th partial trace of the joint output state $\rho^n$ at the eavesdropper, then
\begin{IEEEeqnarray}{rCl}
\delta & \ge & D\left(\rho^n \left\| \rho(0)^{\otimes n} \right.\right) \\
& = & \mathsf{tr} \left[ \rho^n \log \rho^n - \rho^n \log \rho(0)^{\otimes n}\right]\\
& = & - H\left(\rho^n\right) - \mathsf{tr} \left[\rho^n \log \rho(0)^{\otimes n}\right]\\
& = & - H\left(\rho^n\right) - \sum_{i=1}^n \mathsf{tr} \left[ \rho^n \left(I^{\otimes (i-1)} \otimes \log \rho(0) \otimes I^{\otimes(n-i)} \right) \right]\nonumber\\
\ \\
& = & - H\left(\rho^n\right) - \sum_{i=1}^n \mathsf{tr} \left[\rho_i \log \rho(0) \label{eq:rhoi}\right]\\
& \ge & - \sum_{i=1}^n H\left(\rho_i\right) - \sum_{i=1}^n \mathsf{tr} \left[\rho_i \log \rho(0) \right] \\
& = & \sum_{i=1}^n D\left(\rho_i\left\| \rho(0)\right.\right)\\
& \ge & n D(\bar{\rho} \| \rho(0)), \label{eq:converse2}
\end{IEEEeqnarray}
where the last step follows from the convexity of relative entropy.
Here, $\bar{\rho}$ denotes the average output state at the eavesdropper
\begin{equation}
	\bar{\rho} \triangleq \frac{1}{n}\sum_{i=1}^n \rho_i,
\end{equation}
which clearly can also be written as
\begin{equation}
	\bar{\rho} = \sum_{x} \overline{P}(x) \rho(x). \label{eq:converse30}
\end{equation}

Combining \eqref{eq:converse1}, \eqref{eq:converse2}, and \eqref{eq:converse30}, and noting that $\epsilon_n$ must tend to zero as $n$ tends to infinity, we obtain the following:
\begin{equation}\label{eq:IPW}
	L \le \liminf_{n\to\infty} \sqrt{\frac{n}{\delta}}\, \chi(P_n)
\end{equation}
for some $\{P_n, n=1,2,\ldots\}$ subject to the condition that their corresponding output states $\{\rho_ n\}$ at the eavesdropper must satisfy
\begin{equation}\label{eq:Dn}
	D\left(\left.\rho_n\right\|\rho(0) \right) \le \frac{\delta}{n},\quad n=1,2,\ldots.
\end{equation}

Next note that \eqref{eq:Dn} requires 
\begin{equation}\label{eq:rhonto0}
	\lim_{n\to\infty} \rho_n = \rho(0).
\end{equation}
Recall the assumption in Theorem~\ref{thm:main} that no distribution $P$ satisfying \eqref{eq:mixture} can be found. In this case, one can verify that there is no loss of optimality in discarding all $x$ such that $\sigma(x)=\sigma(0)$. After this is done, the only $P$ on the remaining input symbols that satisfies \eqref{eq:mixture1} alone is $P(0)=1$. Thus, to satisfy \eqref{eq:rhonto0}, one needs
\begin{equation}
	\lim_{n\to\infty} P_n(0) = 1.
\end{equation}
Let 
\begin{equation}
	\alpha_n \triangleq 1- P_n(0),
\end{equation}
let $\tilde{P}_n$ denote the distribution given by
\begin{equation}\label{eq:Pnalpha}
	\tilde{P}_n (x) = \frac{P_n(x)}{\alpha_n},\quad x\neq0,
\end{equation}
and let $\tilde{\rho}_n$ denote the state at the eavesdropper induced by $\tilde{P}_n$. Then
\begin{equation}\label{eq:rhonalpha}
	\rho_n = (1-\alpha_n) \rho(0) + \alpha_n \tilde{\rho}_n.
\end{equation}
With \eqref{eq:rhonalpha}, we can fix $\tilde{\rho}_n$ and compute the first and second derivatives of the left-hand side of \eqref{eq:Dn} with respect to $\alpha_n$, using, e.g., \cite[Theorem 11.9]{petz08} and \cite[(7)]{audenaerteisert11}. At $\alpha_n=0$, the first derivative equals zero, and the second derivative is $\eta(\tilde{\rho}_n\| \rho(0))$, yielding (see Lemma~\ref{lem:eta} in the Appendix for details)
\begin{equation}\label{eq:chisquare}
	D\left(\left.\rho_n\right\|\rho(0) \right) = \frac{\alpha_n^2}{2} \eta(\tilde{\rho}_n\| \rho(0)) + o(\alpha_n^2).
\end{equation}
Combined with \eqref{eq:Dn}, this implies
\begin{equation}\label{eq:alphan}
	\alpha_n \le \sqrt{\frac{\delta}{n}} \cdot \frac{1}{\sqrt{\frac{1}{2}\eta(\tilde{\rho}_n\| \rho(0))}} + o(n^{-1/2}).
\end{equation}
On the other hand, by computing the first derivative of the Holevo information $\chi(P_n)$ with respect to $\alpha_n$ (with $\tilde{P}_n$ fixed) we have
\begin{equation}\label{eq:1stder}
\chi(P_n) = \alpha_n \sum_{x\neq 0} \tilde{P}_n(x) D\left( \sigma(x) \left\| \sigma(0)\right.\right) + o(\alpha_n).
\end{equation}
Combining \eqref{eq:IPW}, \eqref{eq:alphan}, and \eqref{eq:1stder}, and recalling that $\alpha_n$ must tend to zero as $n$ tends to infinity, prove the converse part of Theorem~\ref{thm:main}.

\subsection{Achievability}

Our achievability proof is a slight generalization of that in~\cite{sheikholeslami16}. 

Let $\tilde{P}^*$ be the distribution that achieves the maximum in \eqref{eq:main}, and let $\tilde{\rho}^*$ be its corresponding output state at the eavesdropper. Choose the sequence $\alpha_n$, $n=1,2,\ldots$, as
\begin{equation}
\alpha_n = (1-\beta) \sqrt{\frac{\delta}{n}} \cdot \frac{1}{\sqrt{\frac{1}{2} \eta(\tilde{\rho}^*\| \rho(0) )}},
\end{equation}
where $\beta$ is any constant in $(0,1)$. For every $n$, let $P_n$ be given by
\begin{subequations} 
\begin{IEEEeqnarray}{rCl}
	P_n(0) & = & 1-\alpha_n\\
	P_n(x) & = & \alpha_n \tilde{P}^*(x),\quad x\neq 0.
\end{IEEEeqnarray}
\end{subequations}
We randomly generate a codebook (whose size we choose later) by picking every input symbol in every codeword independently and identically distributed (i.i.d.) according to $P_n$.
Using the same approximation as \eqref{eq:chisquare}, we can check that, for large enough $n$, our random codebook satisfies~\eqref{eq:LPD}. Next, by slightly modifying the proof techniques of \cite{hayashinagaoka03}, we obtain a sufficient condition on the sizes of these codebooks for the average probability of a decoding error to tend to zero as $n$ tends to infinity. This condition yields the lower bound
\begin{IEEEeqnarray}{rCl}
L & \ge & \sup \Bigg\{ a \Bigg| \lim_{n\to\infty} \sum_{x^n} \left( \prod_{i=1}^n P_n(x_i)\right)\nonumber\\
	& & ~~~~\cdot \mathsf{tr} \left[ \rho^n(x^n) \left\bra  \widehat{\rho^n}(x^n) - e^{\sqrt{n\delta} a} \rho^n>0\right\ket\right]=1\Bigg\},\IEEEeqnarraynumspace \label{eq:hayashinagaoka}
\end{IEEEeqnarray}
where
\begin{IEEEeqnarray}{rCl}
	&& \rho^n(x^n)  = \rho(x_1)\otimes\cdots\otimes\rho(x_n)\\
	&& \rho^n  = \sum_{x^n} \left(\prod_{i=1}^n P_n(x_i) \right) \rho^n(x^n) = \left(\sum_x P_n(x)\rho(x)\right)^{\otimes n}; \nonumber\\ \ 
\end{IEEEeqnarray}
$\widehat{\rho^n}(x^n)$ is the ``pinching'' of $\rho^n(x^n)$ with respect to $\rho^n$ as defined in \cite{ogawahayashi04}; and $\bra A>0 \ket$ denotes the projector onto the subspace on which the eigenvalues of the operator $A$ are positive. In proving \eqref{eq:hayashinagaoka} we make two main modifications to the original proof techniques of \cite{hayashinagaoka03}. The first modification is to normalize the throughput by $\sqrt{n\delta}$ instead of $n$. The second modification is, instead of the optimal hypothesis test between $\rho^n(x^n)$ and $\rho^n$, which corresponds to a projector of the form
$$ \left\bra \rho^n(x^n) - e^{\sqrt{n\delta} a} \rho^n>0\right\ket,$$
we use a suboptimal test by replacing $\rho^n(x^n)$ with $\widehat{\rho^n}(x^n)$ in this projector. The details of the proof of \eqref{eq:hayashinagaoka} are omitted.

For any $\gamma\in(0,1)$, define the set
\begin{equation}
	\mathcal{A}_n (\gamma) \triangleq \left\{ x^n \left| Q_{x^n}(x) \ge (1-\gamma) P_n(x)  \textnormal{ for all }x\neq 0\right.\right\},
\end{equation}
where $Q_{x^n}$ denotes the ``type'' of the vector $x^n$ \cite{csiszarkorner81}. We shall see that, as $n$ tends to infinity, the probability that $X^n\in\mathcal{A}_n(\gamma)$ tends to one. 
To this end, fix any $x\neq 0$ with $\tilde{P}(x)>0$. Using the Chernoff bound we obtain
\begin{IEEEeqnarray}{rCl}
 \lefteqn{ \mathsf{Pr} \left[ Q_{X^n}(x) < (1-\gamma) P_n(x) \right]}~~~~~~~~~~\nonumber\\
 & \le & \exp \left( - \frac{\gamma^2}{2}\cdot n P_n(x) \right)\\
 & = & \exp \left(-\frac{\gamma^2(1-\beta)\sqrt{\delta} \tilde{P}^*(x)}{\sqrt{2 \eta(\tilde{\rho}^*\|\rho(0))}}\cdot\sqrt{n}\right),\IEEEeqnarraynumspace
\end{IEEEeqnarray}
which tends to zero as $n$ tends to infinity. For $x\neq 0$ and $\tilde{P}(x)=0$, we always have
\begin{equation}
	\mathsf{Pr} \left[ Q_{X^n}(x) < (1-\gamma) P_n(x) \right] = 0.
\end{equation}
Using the union bound and the fact that $\mathcal{X}$ is finite, we then obtain
\begin{equation}
	\lim_{n\to\infty} \mathsf{Pr} \left[ X^n \in \mathcal{A}_n \right] = 1.
\end{equation}

Next we return to \eqref{eq:hayashinagaoka}. Using the method of \cite[Theorem 2]{ogawahayashi04} we have the following: for any parameter $s\in(0,1)$,
\begin{IEEEeqnarray}{rCl}
 \lefteqn{\mathsf{tr} \left[ \rho^n(x^n) \left\bra  \widehat{\rho^n}(x^n) - e^{\sqrt{n\delta} a} \rho^n>0\right\ket\right]}~~~~~~~~~~~~~~~~\nonumber\\ & \ge & 1- (n+1)^k e^{-\psi_n(s)+\sqrt{n\delta} a s},\label{eq:ogawahayashi}
\end{IEEEeqnarray}
where 
\begin{equation}
	\psi_n(s) \triangleq - \log \mathsf{tr} \left[ \rho^n(x^n) (\rho^n)^\frac{s}{2} (\rho^n(x^n))^{-s}(\rho^n)^\frac{s}{2} \right].
\end{equation}
Note that both $\rho^n(x^n)$ and $\rho^n$ are product states, so
\begin{IEEEeqnarray}{rCl}
	\psi_n(s) & = & - \sum_{i=1}^n \log \mathsf{tr} \left[ \rho(x_i) \rho_n^{\frac{s}{2}} \rho(x_i)^{-s} \rho_n^{\frac{s}{2}}\right]\\
	& = & - n \sum_x Q_{x^n} (x) \log \mathsf{tr} \left[ \rho(x) \rho_n^{\frac{s}{2}} \rho(x)^{-s} \rho_n^{\frac{s}{2}} \right]. \label{eq:57}
\end{IEEEeqnarray}
For every $x$, the logarithm term on the right-hand side of \eqref{eq:57} equals zero at $s=0$, and its derivative is uniformly continuous and equals $-D \left(\rho(x)\|\rho_n\right)$ at $s=0$ \cite{ogawahayashi04,sheikholeslami16}. Further note that $\rho_n\to\rho(0)$ as $n\to\infty$. Hence, for any $\theta\in(0,1)$, there exists some $q>0$ such that, for all $s\in(0,q)$, for large enough $n$, and for every $x\neq 0$,
\begin{equation}
	- \log \mathsf{tr} \left[ \rho(x) \rho_n^{\frac{s}{2}} \rho(x)^{-s} \rho_n^{\frac{s}{2}} \right] \ge s (1-\theta)D(\rho(x)\|\rho(0)),
\end{equation}
while
\begin{equation}
	- \log \mathsf{tr} \left[ \rho(0) \rho_n^{\frac{s}{2}} \rho(0)^{-s} \rho_n^{\frac{s}{2}} \right] \ge 0.
\end{equation}
Thus we have, for the above parameters,
\begin{IEEEeqnarray}{rCl}
	\frac{\psi_n(s)}{s} & \ge & (1-\theta) n\sum_{x\neq 0} Q_{x^n}(x) D(\rho(x)\| \rho(0)).
\end{IEEEeqnarray}
For every $x^n \in\mathcal{A}_n$, we further have
\begin{IEEEeqnarray}{rCl}
\frac{\psi_n(s)}{s} & \ge & (1-\theta)(1-\gamma) n \sum_{x\neq 0} P_n(x) D(\rho(x)\| \rho(0))\\
& = & (1-\theta)(1-\gamma) \alpha_n n \sum_{x\neq 0} \tilde{P}^*(x) D(\rho(x)\| \rho(0)) \IEEEeqnarraynumspace\\
& = & (1-\theta)(1-\gamma)(1-\beta) \sqrt{n\delta} \nonumber\\
& & ~~~~~~~~~\cdot \frac{\sum_{x\neq 0} \tilde{P}^*(x) D(\rho(x)\| \rho(0))}{\sqrt{\frac{1}{2} \eta(\tilde{\rho}^*\| \rho(0))}}\\
& = & \sqrt{n\delta}\, \hat{a},\label{eq:63}
\end{IEEEeqnarray}
where for the last step we defined
\begin{equation}
\hat{a} \triangleq (1-\theta)(1-\gamma)(1-\beta) \frac{\sum_{x\neq 0} \tilde{P}^*(x) D(\rho(x)\| \rho(0))}{\sqrt{\frac{1}{2} \eta(\tilde{\rho}^*\| \rho(0))}}.
\end{equation}
Plugging \eqref{eq:63} into \eqref{eq:ogawahayashi} we have that, for every $x^n\in\mathcal{A}_n$,
\begin{IEEEeqnarray}{rCl}
	\lefteqn{\mathsf{tr} \left[ \rho^n(x^n) \left\bra  \widehat{\rho^n}(x^n) - e^{\sqrt{n\delta} a} \rho^n>0\right\ket\right]}~~~~~~~~~~~~~~~~\nonumber\\
	& \ge & 1 - (n+1)^k e^{- \sqrt{n\delta}(\hat{a}-a) s}.
\end{IEEEeqnarray}
Recall that, as $n$ tends to infinity, the probability that $X^n\in\mathcal{A}_n$ tends to one. Thus for every $a<\hat{a}$,
\begin{IEEEeqnarray}{rCl}
\lefteqn{\lim_{n\to\infty} \sum_{x^n} \left(\prod_{i=1}^n P_n(x_i) \right)}~~~ \nonumber\\
	& & ~~~~{}\cdot \mathsf{tr} \left[ \rho^n(x^n) \left\bra  \widehat{\rho^n}(x^n) - e^{\sqrt{n\delta} a} \rho^n>0\right\ket\right]\nonumber\\
& \ge  & \lim_{n\to\infty} \mathsf{Pr} \left[ X^n \in \mathcal{A}_n \right] \cdot \left(1 - (n+1)^k e^{- \sqrt{n\delta}(\hat{a}-a) s} \right)\IEEEeqnarraynumspace \\
& = & 1. \label{eq:equalsone}
\end{IEEEeqnarray}
Finally, since $\theta$, $\gamma$, and $\beta$ can all be chosen to be arbitrarily small, \eqref{eq:equalsone} holds for all
\begin{equation}
a <  \frac{\sum_{x\neq 0} \tilde{P}^*(x) D(\rho(x)\| \rho(0))}{\sqrt{\frac{1}{2} \eta(\tilde{\rho}^*\| \rho(0))}}.
\end{equation}
Combining this fact with \eqref{eq:hayashinagaoka} proves the achievability part of Theorem~\ref{thm:main}.

\section{Concluding Remarks}\label{sec:conclusion}

We have seen that previous analysis of covert communication over classical channels can be combined with quantum information theory to derive analogous results on classical-quantum channels. 

Our converse proof does not seem easily extendable to covert communication of classical information over memoryless quantum (not classical-quantum) channels, for which it is known that entangled input states may result in ``super additivity'' of Holevo information \cite{hastings09}. Determining whether the square-root law also holds for quantum channels may be an interesting problem for future work.


\bibliographystyle{hieeetr}           
\bibliography{/Volumes/Data/wang/Library/texmf/tex/bibtex/header_short,/Volumes/Data/wang/Library/texmf/tex/bibtex/bibliofile}

\appendix

\begin{lemma}\label{lem:eta}
Let $A$ and $B$ be density operators acting on the same Hilbert space, then
\begin{equation}
\lim_{t\downarrow 0} \frac{D \bigl( t A + \bigl. (1-t) B  \bigr\| B\bigr)}{t^2} = \frac{1}{2} \eta(A\|B),
\end{equation}
where $\eta(\cdot\|\cdot)$ is defined in \eqref{eq:defeta}.
\end{lemma}

\begin{proof}
We compute the first and second derivatives of the function $t\mapsto D \bigl( t A + \bigl. (1-t) B  \bigr\| B\bigr)$, $t\in[0,1]$, at $t=0$. Write
\begin{IEEEeqnarray}{rCl}
\lefteqn{D \bigl( t A + \bigl. (1-t) B  \bigr\| B\bigr)}~~~~~~~\nonumber\\
 & = & \tr \left[\bigl( t A + \bigl. (1-t) B \bigr) \log \bigl( t A + \bigl. (1-t) B \bigr)\right] \nonumber\\* & & {}- \tr\left[\bigl( t A + \bigl. (1-t) B \bigr) \log B \right].\label{eq:eta69}
\end{IEEEeqnarray}
The term inside the first trace on the right-hand side of \eqref{eq:eta69} can be written as a matrix function $f\bigl( t A + \bigl. (1-t) B \bigr)$, where $f(a)=a\log a$ for $a>0$, so its derivative with respect to $t$ can be computed using \cite[Theorem 11.9]{petz08}. The second term on the right-hand side of \eqref{eq:eta69} is linear in $t$. We hence obtain 
\begin{IEEEeqnarray}{rCl}
\lefteqn{\frac{\d }{\d t} D \bigl( t A + \bigl. (1-t) B  \bigr\| B\bigr)}~~~~~~~~\nonumber\\* & = & \tr\left[ (A-B) \left(\log\bigl( tA + (1-t)B \bigr) + 1 \right)\right] \nonumber\\& & {} - \tr \left[ (A-B) \log B \right]\\
& = & \tr\left[ (A-B) \log\bigl( (tA + (1-t)B\bigr)\right] \nonumber\\ & & {}- \tr \left[ (A-B) \log B \right]. \ \label{eq:1st}
\end{IEEEeqnarray}
From \eqref{eq:1st} we immediately have
\begin{equation}
\left.\frac{\d }{\d t} D \bigl( t A + \bigl. (1-t) B  \bigr\| B\bigr) \right|_{t=0} = 0. \label{eq:eta1st}
\end{equation}
We continue from \eqref{eq:1st} to compute the second derivative. Using \cite[(7)]{audenaerteisert11} we have
\begin{IEEEeqnarray}{rCl}
\lefteqn{\left.\frac{\d^2}{\d t^2} D \bigl( t A + \bigl. (1-t) B  \bigr\| B\bigr)\right|_{t=0}}\nonumber\\
 & = &  \tr \left[ \left.(A-B) \left\{\frac{\d}{\d t}\log \bigl( (tA + (1-t)B \bigr)\right|_{t=0}\right\}\right]\IEEEeqnarraynumspace\\
& = & \tr \left[ (A-B) \int_0^\infty (B+s)^{-1} (A-B) (B+s)^{-1} \d s \right]\IEEEeqnarraynumspace\\
& = & \tr \left[ \int_0^\infty A (B+s)^{-1} A (B+s)^{-1} \d s\right] -1,\IEEEyesnumber \label{eq:eta2nd}
\end{IEEEeqnarray}
where the last step follows by standard matrix manipulations and by the fact
\begin{equation}
\int_0^\infty (B+s)^{-1} B (B+s)^{-1} \d s = I.
\end{equation}
Recalling \eqref{eq:defeta}, the right-hand side of \eqref{eq:eta2nd} is $\eta(A\|B)$. Hence
combining \eqref{eq:eta1st} and \eqref{eq:eta2nd} completes the proof.
\end{proof}

\section*{Acknowlegement}

The author thanks Koenraad Audenaert and Mil\'an Mosonyi for their help in proving Lemma~\ref{lem:eta}.

\end{document}